\documentclass[twoside,fleqn]{ActaStyle}
\usepackage[dvips]{graphics}
\usepackage{times,cite}
\usepackage{graphicx}
\usepackage{floatflt}
\usepackage{color}

\def\authorheading{J. Speltz}
\def\shorttitle{Energy Dependence of Strangeness Production At {\sc rhic}}
	
\begin{document}
\pagerange{1}{5}   

\title{Energy Dependence Systematics of Strange and Multi-Strange Particle Production}

\author{J.~Speltz\email{jeff.speltz@ires.in2p3.fr}$^*$ for the STAR Collaboration}{$^*$ Institut de Recherches Subatomiques, Strasbourg, France}

\abstract{We present results on systematic measurements of strange and multi-strange particles with the {\sc star} detector for center of mass energies per nucleon pair ($\sqrt{s_{NN}}$) of 62.4 and 200 GeV in ultra-relativistic Au+Au collisions at {\sc rhic}. We use these results to characterize the chemical and kinetic freeze-out properties of the fireball produced by the collision. This is done by comparison to statistical and hydrodynamical models as well as parameterization. 
We emphasize particularly the energy dependence of these features comparing measurements obtained at {\sc rhic} with different energies but also at the lower energies available at the {\sc sps}.} 

 
\section{Introduction}

The measurement of strangeness in relativistic heavy ion collisions has been suggested to be valuable for gaining insight in the created system, as strange valence quarks present in the final state of the collision are not existant in the incoming beams. Studying the chemical or kinetic freeze-out properties of strangeness in general and compared to non strange particles in particular may then reveal different aspects and characteristics of the fireball they originate from. With the results from the 62.4 GeV energy that stands between the top {\sc rhic} and the {\sc sps} energies a more complete picture of the excitation function of these properties can be accessed. 

The data used for the presented results are from the {\sc star} experiment for Au+Au collisions and were obtained with the main tracking device of {\sc star} \cite{ref:STAR}, the Time Projection Chamber ({\sc tpc}). The 200 GeV strangeness results are from the RUN II data \cite{ref:SQM_estienne}, while the 62.4 GeV results are from the RUN IV data \cite{ref:SQM_speltz,ref:SQM_takahashi}.

\section{Chemical Properties}

The chemical freeze-out marks the end of all inelastic interactions and fixes the relative aboundance of the different particle species. Statistical models compute these production rates using a limited set of parameters and assuming the freeze-out occurs from a statistically equilibrated matter. Hence one can consider that these parameters characterize the chemical properties of the system. These parameters are the temperature at freeze-out $T_{ch}$, the chemical potentials of light quarks $\mu_q$ ($q=u,d$) and strange quarks $\mu_s$ as well as the strange quark phase-space occupancy $\gamma_S$. For a given collision energy and system, the temperature, potentials and occupancy factor are adjusted to the measured ``stable'' particles ratios (no resonance included).

\begin{figure}[t]
\begin{minipage}[c]{0.47\textwidth}
\begin{center}
\includegraphics[width=6.5cm]{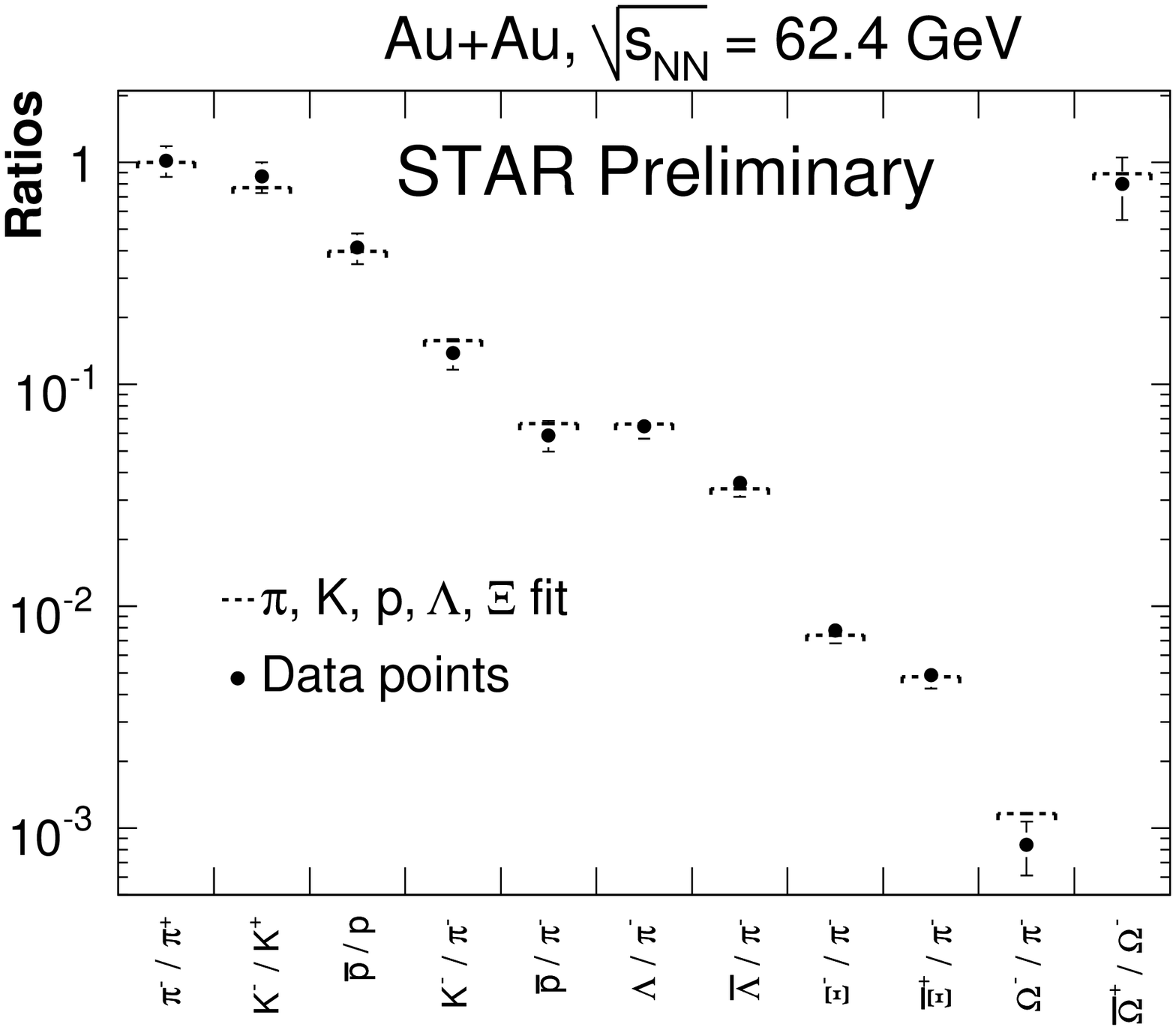}
\vspace{-0.8cm}
\caption{``Stable'' particle ratios as obtained by measurements for most central events at $\sqrt{s_{NN}}=62.4$~GeV (circles) together with statistical model predictions using \cite{ref:Kaneta_Xu} (lines). The experimental data include statistical and systematic errors.}\label{fig:Stat_Ratios}
\end{center}
\end{minipage}
\begin{minipage}[c]{0.06\textwidth}
\hspace{0.5cm}
\end{minipage}
\begin{minipage}[c]{0.47\textwidth}
\begin{center}
\includegraphics[width=6.5cm]{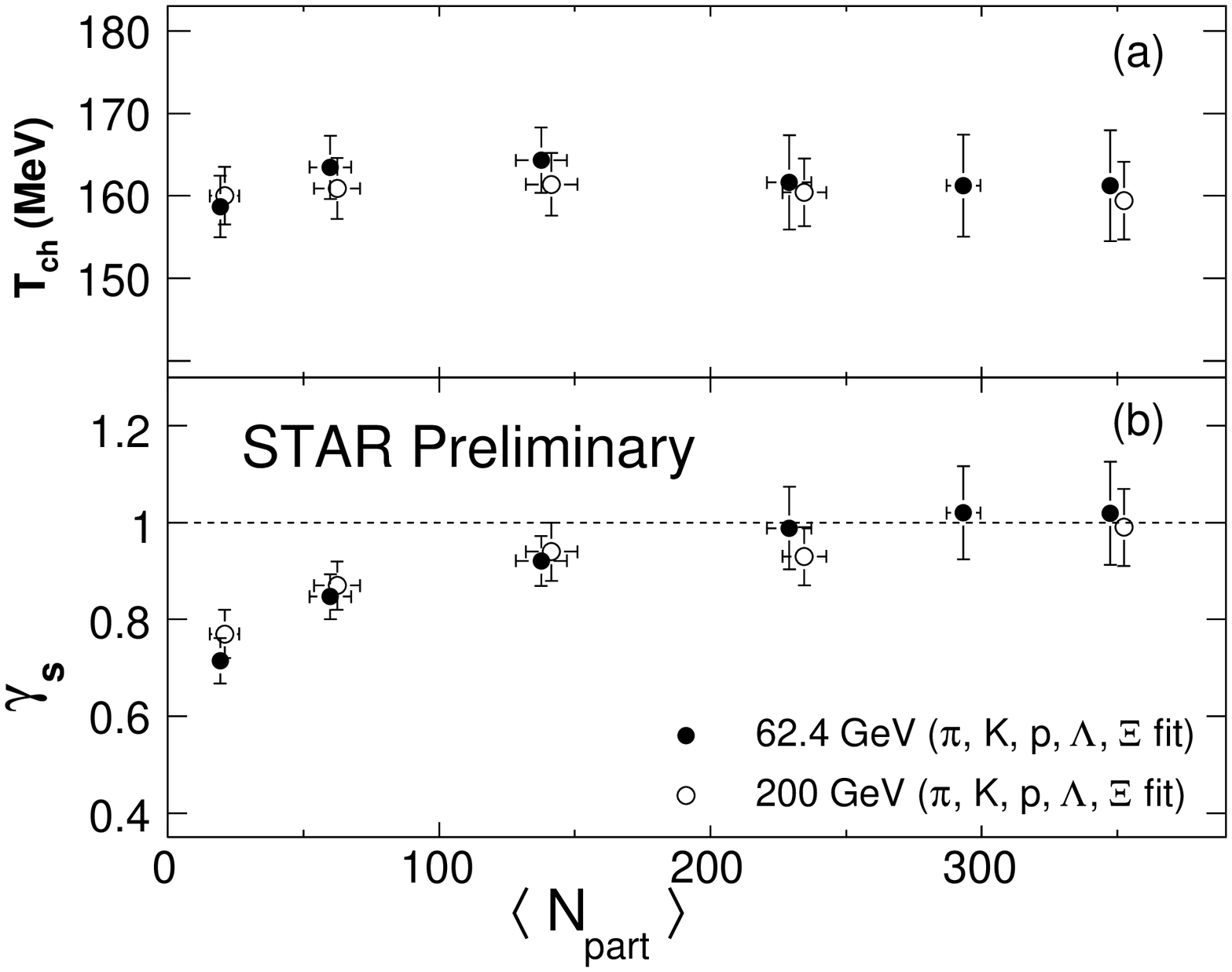}
\vspace{-0.8cm}
\caption{Centrality dependence of (a) chemical freeze-out temperature $T_{ch}$ and (b) strangeness saturation factor $\gamma_S$ for 62.4 GeV (closed circles) and 200 GeV (open circles). Statistical and systematic errors are included.}\label{fig:Stat_Tch_gammaS}
\end{center}
\end{minipage}
\vspace{-0.3cm}
\end{figure}

{\parindent0pt\begin{floatingfigure}[l]{5.5cm}
\begin{center}
\vspace{-0.4cm}
\includegraphics[width=5.5cm]{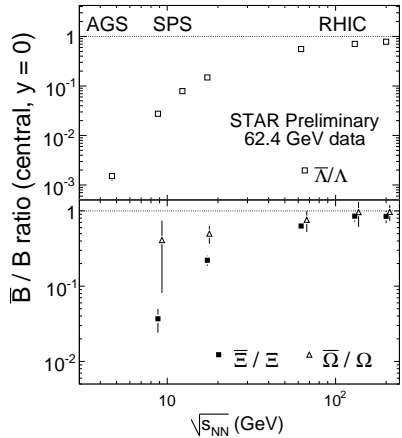}
\vspace{-0.8cm}
\caption{Central, mid-rapidity strange anti-baryon to baryon ratios as a function of $\sqrt{s_{NN}}$ \cite{ref:Caines_Antinori_Adler_Anticic_Adams}.}\label{fig:BbarBratio}
\vspace{-0.2cm}
\end{center}
\end{floatingfigure}
}

Fig.\ref{fig:Stat_Ratios} shows the different particle ratios as obtained by measurements for central collisions at $\sqrt{s_{NN}}=62.4$~GeV together with a statistical model fit \cite{ref:Kaneta_Xu} giving $T_{ch}=161\pm 7$~MeV and $\mu_B = 87 \pm 13$~MeV. The equivalent figure for 200 GeV can be found in \cite{ref:Barannikova}. When comparing to the results from {\sc sps} \cite{ref:PBM_1999}, it can be assessed that statistical models have been very successful in predicting 
``stable'' particle ratios, including multi-strange baryons, on a large range of energies. 

Fig.\ref{fig:Stat_Tch_gammaS} shows the evolution of $T_{ch}$ and $\gamma_S$ with respect to the mean number of participants $\langle N_{part} \rangle$ as determined by a Glauber model calculation \cite{ref:Glauber}, for 62.4 and 200 GeV. Firstly, it can be seen that $T_{ch}$ is almost constant over the total centrality range and for both energies. Its value of approximately $160$~MeV is close to the predicted LQCD phase transition temperature of around $170$~MeV \cite{ref:Karsch_Redlich_Tawfik}. Secondly, $\gamma_S$ shows a centrality dependence rising from most peripheral collisions and reaching unity for most central collisions. Here again the results from 62.4 GeV and 200 GeV are very similar and may reveal equilibration of strangeness for most central collisions at {\sc rhic} energies, with contrast to {\sc sps} ones where also a centrality dependence for $\gamma_S$ has been manifested, but the value for most central events seems to be less than unity \cite{ref:Cleymans}.

On Fig.\ref{fig:BbarBratio} the evolution of the strange anti-baryon to baryon ratio for central collisions at mid-rapidity with collision energy, from {\sc ags} over {\sc sps} to {\sc rhic}, is presented. With the newly added 62.4 GeV value these ratios yield a smooth increase in going from a transport preponderant regime at the lower energy to reaching a baryon free regime where quark pair production is dominant at the highest {\sc rhic} energy.
 
\section{Dynamical Properties}

While chemical properties can be derived from transverse momentum ($p_T$) integrated values, $p_T$ dependent distributions, such as the particle spectra and differential elliptic flow \cite{ref:Ollitrault} measurement are needed to gather information on dynamical characteristics. These are among others the kinetic freeze-out temperature $T_{kin}$ at which all elastic interactions end, but also the flow velocity (elliptic and radial) developed during the expansion of the medium. 

\begin{figure}[h]
\begin{minipage}[c]{0.485\textwidth}
\begin{center}
\vspace{-0.2cm}
\hspace{-0.4cm}\includegraphics[width=6.9cm]{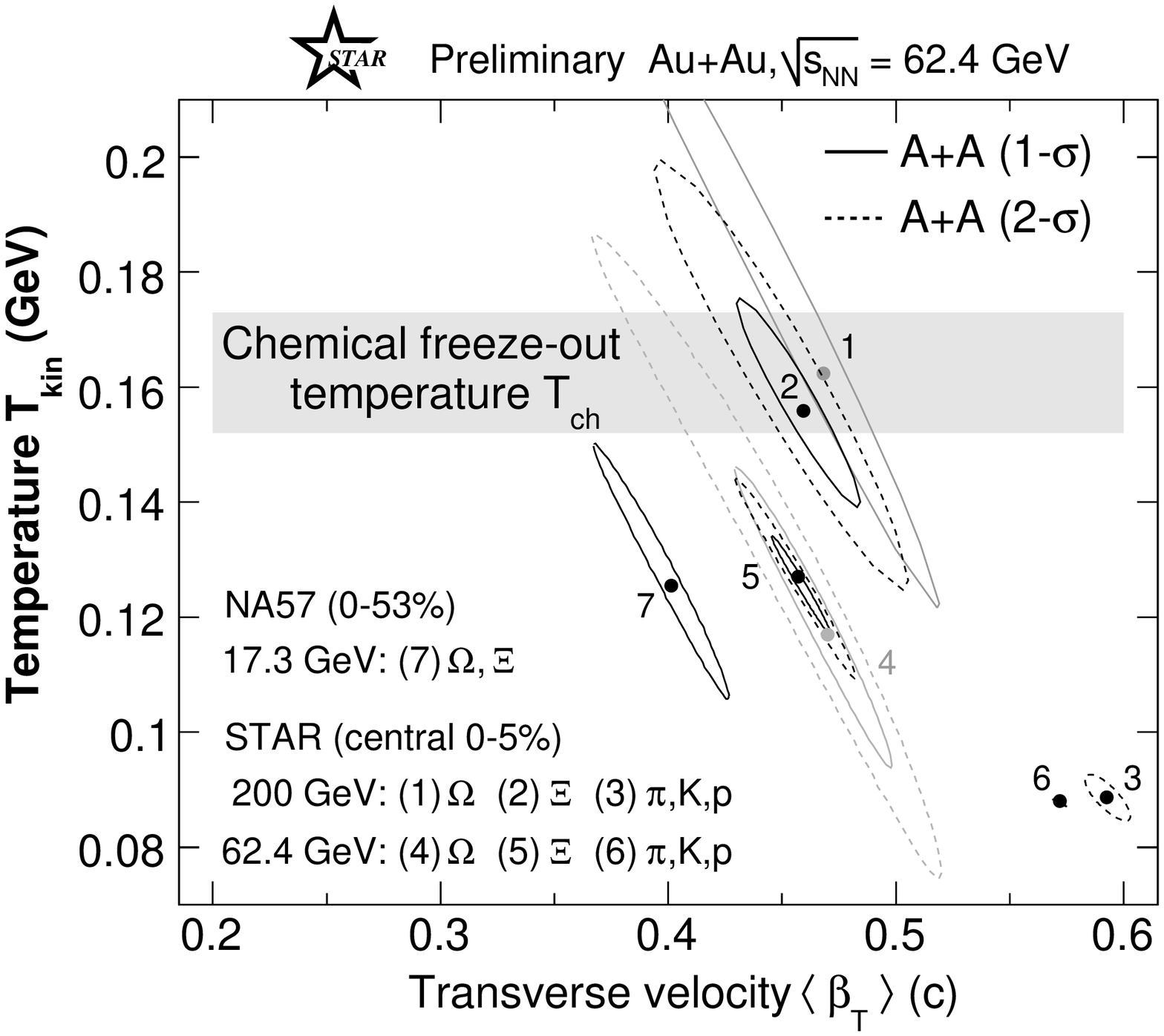}
\vspace{-0.4cm}
\caption{ One and two $\sigma$ contours in $T_{kin}-\langle \beta_{T} \rangle$ space obtained by blast-wave parameterization for most central collisions at different energies \cite{ref:SQM_estienne,ref:Alt} and for different particle species. }\label{fig:BWContours}
\end{center}
\end{minipage}
\begin{minipage}[c]{0.03\textwidth}
\hspace{0.2cm}
\end{minipage}
\begin{minipage}[c]{0.485\textwidth}
\begin{center}
\includegraphics[width=6.cm]{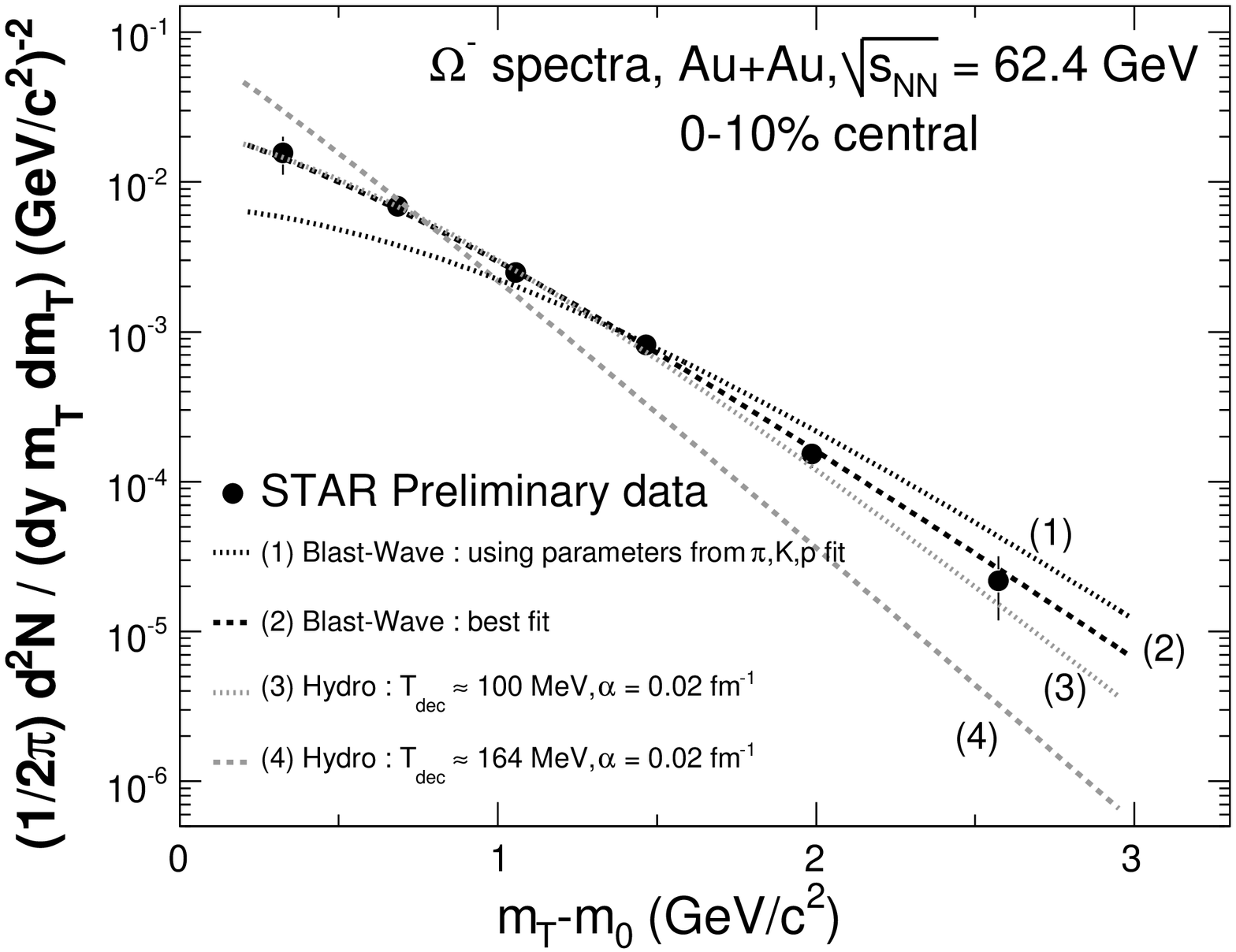}
\vspace{-0.3cm}
\caption{Transverse mass $\left( m_T = \sqrt{p_T^2+m_0^2} \right)$ $\Omega^-$ spectra obtained by measurement at $\sqrt{s_{NN}}=62.4$~GeV (black points) and comparison to blast-wave parameterization (black lines) and hydrodynamical calculation \cite{ref:Kolb_Sollfrank_Heinz} (grey lines). Errors on the spectra are statistical only. See text for more details.}\label{fig:SpectraFit}
\end{center}
\end{minipage}
\end{figure}

It has been tried to infer information on these parameters by using either an ideal hydrodynamical description \cite{ref:Kolb_Sollfrank_Heinz} which gives a prediction on the spectra and elliptic flow or a hydrodynamical inspired blast-wave parameterization \cite{ref:Retiere_Lisa,ref:Schnedermann} which permits the extraction of parameters by doing a direct fit on the spectra. In the case of the blast-wave these parameters are $T_{kin}$, the mean tranverse flow velocity $\langle \beta_T \rangle$, but also the velocity radial profile $n$. In the used hydrodynamical model, in addition to the decoupling temperature $T_{dec}$ which is the average temperature on the decoupling surface, a variable that parameterizes the initial transverse boost at thermalization (noted $\alpha$) was used. 

Fig.\ref{fig:BWContours} shows confidence-level contours in $T_{kin}-\langle \beta_{T} \rangle$ space as obtained by blast-wave fits on the spectra of different particle species and different energies. The 62.4~GeV and 200~GeV contours reveal a clear discrepancy of the transverse flow of the light particles ($\pi$, K and p) and the multi-strange baryons ($\Xi$ and $\Omega$). Such a disagreement may also be seen on $T_{kin}$ at 200~GeV, but is not as obvious at 62.4~GeV. It has been argued at 200~GeV \cite{ref:SQM_estienne} that $T_{kin}$ is compatible with $T_{ch}$ for multi-strange baryons and thus leading to limited interactions of these particles after chemical freeze-out. At 62.4~GeV $T_{ch}$ is comparable to the 200~GeV value whereas $T_{kin}$ obtained by blast-wave fits for $\Xi$ and $\Omega$ is lower as at 200~GeV. This would then lead to the possibility that the multi-strange baryons may continue interacting a little longer after chemical freeze-out at the lower energy than at the higher energy. Additionally it can also be seen that $\langle \beta_T \rangle$ is increasing with increasing energy for the light particles, while there may be an indication for an increase of temperature for the multi-strange baryons when going from 62.4~GeV to 200~GeV, although the values are still compatible within errors.

Fig.\ref{fig:SpectraFit} shows the $\Omega^-$ spectra obtained by measurement at 62.4 GeV together with results from blast-wave parameterization (black lines) and hydrodynamical calculation (grey lines). The black dashed line shows the best fit from blast-wave with the data. This best fit gives T$_{kin} \approx 120 \pm 30$~MeV and $\langle \beta_T \rangle \approx 0.47 \pm 0.05$~c.  These values show a little dependence of the velocity profile which was fixed to $n=1$. The black dotted line gives the spectra obtained by blast-wave when fixing the parameters to the values obtained by a simultaneous fit to the spectra from $\pi^-$, $\pi^+$, K$^-$, K$^+$, p and $\overline{\mbox{p}}$ ($T_{kin}=90$~MeV and $\langle \beta_T \rangle = 0.57$~c). As has already been observed qualitatively on Fig.\ref{fig:BWContours}, the spectral shape of the black dotted line is in clear disagreement with the measurement. The difference in the lower $p_T$ part may partially be attributed to incompatibility in flow, while the deviation in the higher $p_T$ part may be due to discrepancy in freeze-out temperature between the light particles and the $\Omega$. As already indicated in Fig.\ref{fig:BWContours} the former seems to be more pronounced. 

This observation is supported by the spectral shapes obtained by hydrodynamical calculation. The use of a temperature $T_{dec}=100$~MeV (grey dotted line on Fig.\ref{fig:SpectraFit}) and $\alpha=0.02$~fm$^{-1}$ (the use of $\alpha=0$~fm$^{-1}$ reveals a larger deviation from measurement at higher $p_T$) gives a form that is in agreement with the data, while the shape obtained with a higher temperature (grey dashed line) clearly misses that from the data. This has also already been seen at 200 GeV \cite{ref:Kolb_Sollfrank_Heinz}. 

We discuss here only the shape of the spectra and not its normalization as the used hydrodynamical model does not include a strange chemical potential that may have significant influence on the strange particle multiplicity. Therefore the normalization has been scaled in order to reproduce the data. Additionally systematic differences were checked between the hydrodynamical and blast-wave temperature by producing a spectra with hydrodynamical model using a certain temperature ($T_{dec}$) and then performing a blast-wave fit. We obtain a good agreement, however the values obtained by the blast-wave fit are systematically lower than the hydrodynamical temperature with a smaller difference for lower $T_{dec}$ ($3-10$~MeV at $T_{dec} \approx 100$~MeV) than at higher $T_{dec}$ ($15-25$~MeV at $T_{dec} \approx 160$~MeV). 

Finally, we also investigated differential elliptic flow resulting from the azimuthal asymmetry in momentum space. This elliptic flow has been compared to hydrodynamical prediction. The 62.4 GeV results are not shown, but give similar results than at 200 GeV \cite{ref:Oldenburg}. The mass ordering at low-$p_T$ is well reproduced by hydrodynamics as well as the magnitude up to $p_T$ of around 2~GeV/c. This was not the case at the {\sc sps}, where the hydrodynamic limit was not reached \cite{ref:Alt2}.
 
\section{Conclusion}

The high quality and quantity of data accumulated both at {\sc sps} and {\sc rhic} for different systems and at different energies allow for systematic comparison and excitation function of physics observables including strange and even multi-strange particles. 62.4~GeV data give qualitatively comparable results as the 200~GeV measurements in term of chemical and dynamical properties. The hydrodynamical calculation and blast-wave parameterization seem to work as good at the lower {\sc rhic} energy (62.4 GeV) as at 200 GeV and give comparable results for freeze-out temperature. However, the transverse flow obtained by blast-wave for the multi-strange particles is significantly lower than that for the light particles. Finally, although ideal hydrodynamics seem to work quite well, it is still interesting to see how models that are not assuming such an ideal behavior can describe the data, in order to get the entire picture of the properties of the matter created at {\sc rhic}.

\end{document}